\documentclass[aps,prl,preprint,groupedaddress]{revtex4-1}
\usepackage{amsmath,amssymb,ulem}
\usepackage{graphicx,mathtools,xcolor,color,CJK}

\newcommand{\partialderiv}[3][]{\frac{\partial^{#1}#2}{\partial {#3}^{#1}}}

\def\_#1{{\underline{#1}}}

\newcommand{\bp}{\begin{pmatrix}}
\newcommand{\ep}{\end{pmatrix}}

\newcommand{\be}{\begin{equation}}
\newcommand{\ee}{\end{equation}}

\newcommand{\sech} { {\rm sech} \hskip 0.01in}

\newcommand{\mn}{\mathbb{E}}
\newcommand{\boldxi}{\boldsymbol{\xi}}
\newcommand{\boldrho}{\boldsymbol{\rho}}
\newcommand{\Li}{\mathrm{Li}\,}
\newcommand{\sgn}{\mathrm{sgn}\,}
\newcommand{\heff}{{\bf h}_{\mathrm{eff}}}
\newcommand{\hatm}{\hat{\bf m}}
\newcommand{\vpar}{v_{\parallel}}
\newcommand{\vperp}{v_{\perp}}

\begin{document}

\title{Stochastic ejection of nanocontact droplet solitons via drift instability}

\author{Richard O.\ Moore}
\affiliation{Department of Mathematical Sciences, New Jersey Institute of Technology, Newark, New Jersey 07102}
\email{rmoore@njit.edu}

\author{Mark A.\ Hoefer}
\affiliation{Department of Applied Mathematics, University of Colorado, Boulder, Colorado 80302}

\date{\today}

\begin{abstract}
  The magnetic droplet soliton is a large amplitude, coherently
  precessing wave state that exists in ferromagnetic thin films with
  perpendicular magnetic anisotropy.  To effectively sustain a
  droplet, magnetic damping can be locally compensated in a
  nanocontact region that imparts spin-transfer torque; this has been successfully
  deployed in experiment to directly image the droplet and probe its
  dynamics electrically.  However, theory predicts and experiments
  indicate the existence of a drift instability whereby the droplet is
  ejected from the spin-transfer-torque-active region and subsequently decays,
  an effect that may be enhanced or possibly induced by thermal fluctuations.  Using
  soliton perturbation theory and large deviation theory, this work
  determines the soliton ejection rate and the most likely path an
  ejected soliton tracks in the presence of thermal fields.  These
  results lead to an effective lower bound on the stability of
  magnetic droplet solitons in spin-transfer torque nanocontact devices
  operating at finite temperature and point to ways in which droplets
  can be made more robust.
\end{abstract}


\maketitle

\section{Introduction}

The pursuit of fast, scalable nonvolatile storage and new ways to process digital information have motivated much recent study of the formation and control of localized structures in ferromagnetic materials.  These structures are supported by an energetic balance among contributions from material anisotropy, exchange energy, dipole or self-induced field energy, and forcing due to external fields, current-induced spin torque, and magnetic damping.  Some localized structures, such as skyrmions, are created with a topology that cannot continuously deform to the trivial state, lending promise for additional structural stability in the absence of a driving force.  Other structures, such as dissipative solitons, are maintained by a balance between driving and damping terms.  In both cases, structures that are stable under deterministic dynamics are generally only quasi-stable under the inevitable influence of thermal fluctuations or other stochastic phenomena.

In this paper, we study the effects of weak thermal fluctuations on
the magnetic droplet soliton \cite{kosevich_magnetic_1990}, a
large-amplitude, precessing mode that manifests as a balance between
nonlinearity due to uniaxial anisotropy and dispersion due to the
exchange interaction.  Because of the droplet's precessional dynamics,
its practical realization was predicted \cite{hoefer_theory_2010} and
then observed both indirectly via the giant magnetoresistance (GMR)
effect \cite{mohseni_spin_2013,macia_stable_2014} and directly with
x-ray transmission microscopy
\cite{backes_direct_2015,chung_direct_2018} in a spin-transfer torque driven
nanocontact device that imparts a current-induced local torque to compensate uniform
magnetic damping.  Additional experiments
\cite{lendinez_observation_2015,chung_magnetic_2016} used the GMR
effect to measure the nanocontact device's frequency characteristics,
highlighting a low-frequency signal that was attributed to a drift
instability whereby the soliton is expelled from the nanocontact and then
decays, after which a new droplet forms under the nanocontact and the process
repeats \cite{hoefer_theory_2010}.  This drift instability presents a
significant challenge to the robust control of droplet solitons and
the reliable operation of future spintronic devices.  Due to
growing interest in magnetic droplet solitons, experiments
continue in spin torque nanocontacts
\cite{xiao_parametric_2017,hang_generation_2018,chung_direct_2018,jiang_using_2018} and
in spin Hall driven nanoconstrictions \cite{divinskiy_magnetic_2017}.

The existence and dynamics of droplets in nanocontacts have been
successfully described using soliton perturbation theory
\cite{bookman_analytical_2013,bookman_perturbation_2015}, that yields
a system of ordinary differential equations (ODEs) for the droplet's
parameters that we call the modulation equations.  The ODEs' fixed points correspond to sustained droplets,
which, for a certain parameter regime, are unstable and manifest an
increase in the droplet's speed, i.e., the drift instability
\cite{wills_deterministic_2016}.  Their stochastic dynamics under a
thermal random field were explored
theoretically~\cite{wills_deterministic_2016}.  The objective of that
study was to compute the approximate variance in droplet parameters,
including its frequency, linewidth and position relative to the spot
of the nanocontact supporting it.  The Ornstein-Uhlenbeck process
derived from linearizing the stochastic model equations about the
dissipative droplet state was shown to provide statistics that agree
well with simulations of the nonlinear stochastic model and with
micromagnetic simulations performed using open source code MuMax3.
That study also contained the description of the aforementioned drift
instability occurring at critical bias current.

An equally important figure of merit for magnetic droplets in the presence of a thermal field is the probability (alternatively, the rate) of ejection, where the droplet escapes the attractive potential provided by the finite spatial extent of the nanocontact and then decays due to magnetic damping.  In the physically relevant limit of weak thermal effects, exits occur rarely, with a probability or rate determined by large deviation theory.  In this limit, the asymptotic scaling of exit probabilities and rates with vanishing noise strength is provided by a rate function related to the
exit path or paths that minimize an action functional.  This asymptotic scaling provides an order-of-magnitude estimate for the ejection rate, which would otherwise be estimated through sampling using, for example, a software package such as MuMax3~\cite{vansteenkiste_design_2014}.  These simulations must resolve dynamics on gyromagnetic time scales of nanoseconds, while our estimates show that exits occur on time scales of micro- to milliseconds, depending on the operating regime.  Monte Carlo methods to estimate ejection rates or ejection probabilities over finite times are therefore unreasonably expensive.

In this work, we demonstrate that, even in regimes where the droplet fixed point is nominally stable under deterministic dynamics of the modulation ODEs, the proximity of a saddle point associated with the drift instability provides an important mechanism for droplet ejection.  We use a combination of analytical and numerical techniques to compute the action associated with exits through all saddles with a single unstable manifold, including the saddles with nonzero velocity bifurcating through the drift instability~\cite{wills_deterministic_2016} and those associated with zero velocity originating in a saddle-node bifurcation in the precessional frequency~\cite{bookman_analytical_2013}.  Our most important physical result is the determination of the mean time it takes for a deterministically stable droplet to be ejected from the nanocontact due to thermal noise.  Based on physical parameters from recent experiments~\cite{lendinez_effect_2017,lendinez_observation_2015}, we estimate this time to be approximately 50~nanoseconds or a rate of droplet ejection of 20~MHz, which is within an order of magnitude of low-frequency observations from recent experiments.  We estimate that this rate can be manipulated and, importantly, significantly decreased through appropriate choice of operating parameters.

\section{Model}

We study a model for the magnetization ${\bf M}=M_s\hatm$ of the free layer in a thin magnetic film supported from below by a thin nonmagnetic conducting layer followed by a fixed layer with constant magnetization ${\bf M}_p=M_p\hatm_p$, where $M_s$ and $M_p$ are their respective saturation magnetizations.  The fixed layer plays the role of a current polarizer enabling both magnetoresistive detection of the free layer magnetization and switching of the free layer through spin-transfer torque (STT).  The current density is approximately restricted in space to a circular disc nanocontact of radius $R_*$ placed on top of the free layer.
An external field $H_0$ is applied perpendicular to the layers in order to provide control over the fixed layer magnetization and to stabilize the droplet.

From the Landau-Lifshitz equations for the free layer normalized magnetization defined on the plane, $\hatm({\bf x},t):{\mathbb R}^2\times{\mathbb R}\rightarrow{\mathbb S}^2$, we have
\be
\partialderiv\hatm{t} = -\hatm\times \heff -\alpha\hatm\times(\hatm\times \heff) + i H(\rho_*-|{\bf x}|)\hatm\times(\hatm\times\hatm_p)
-\hatm\times{\bf h},
\label{e:LL}
\ee
where $0<\alpha\ll 1$ is the damping parameter, $H(x)$ is the Heaviside function, and $i=I/I_0$ is the STT current nondimensionalized by
\be
I_0 = \frac{4\mu_0M_s(H_k-M_s)e\pi R_*^2\delta}{\hbar\eta},
\ee
with $H_k$ the anisotropy field, $\delta$ the free layer thickness, $\eta$ the spin torque efficiency, and $\rho_*=R_*/L$.  In Eqn.~\ref{e:LL}, time has been nondimensionalized using the precessional time scale $\tau=(|\gamma|\mu_0(H_k-M_s))^{-1}$, where $\gamma$ is the gyromagnetic ratio and $\mu_0$ is vacuum permeability.
In-plane lengths are normalized by $L=\lambda_{\mathrm{ex}}\sqrt{M_s/(H_k-M_s)}$ where $\lambda_{\mathrm{ex}}$ is the exchange length.
The respective influences of the external magnetic field $H_0 = (H_K-M_s)h_0$, the exchange field, and perpendicular anisotropy sufficient to exceed the local demagnetizing field contribution ($H_K>M_s$) are included in
\be
\heff = h_0\hat{\bf z}+\nabla^2\hatm+m_z\hat{\bf z}.
\ee
Finally, thermal effects are modeled using the fluctuation-dissipation principle~\cite{brown_jr_thermal_1963} and are described by the final term in Eqn.~\ref{e:LL}, which is taken to be space-time white noise with
\be
{\mathbb E}({\bf h}({\bf x},t){\bf h}^\dag({\bf x}',t')) =
\beta^2{\bf I}\,\delta({\bf x}-{\bf x}')\delta(t-t'),
\ee
where ${\bf I}$ is the identity matrix, $T$ is temperature, and $\beta^2=T/T_0$, with
\[
T_0 = \frac{\mu_0M_s^2\lambda_{\mathrm{ex}}^2\delta}{2\alpha k_B}.
\]
It is important to note that the stochastic differential equation (SDE) Eqn.~\ref{e:LL} must be interpreted in the sense of Stratonovich to preserve the unit magnitude of $\hatm$ ($|\hatm({\bf x},t)|=1$ for all ${\bf x}\in{\mathbb{R}}^2$, $t\geq 0$).

\begin{figure}
\centerline{\includegraphics[width=0.5\textwidth]{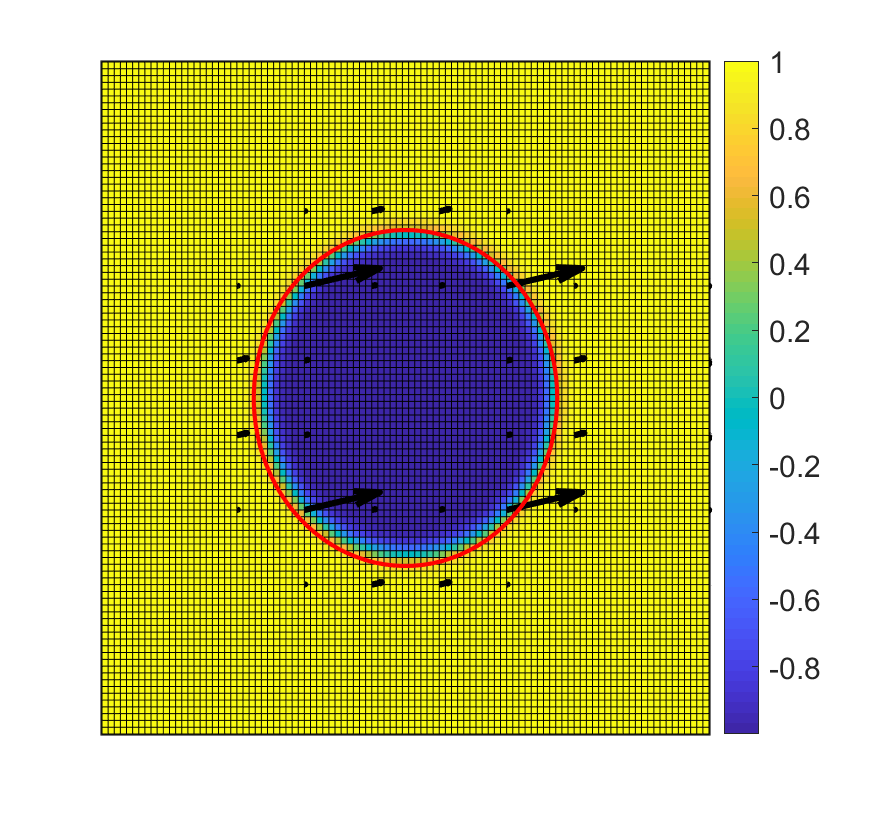}
\includegraphics[width=0.5\textwidth]{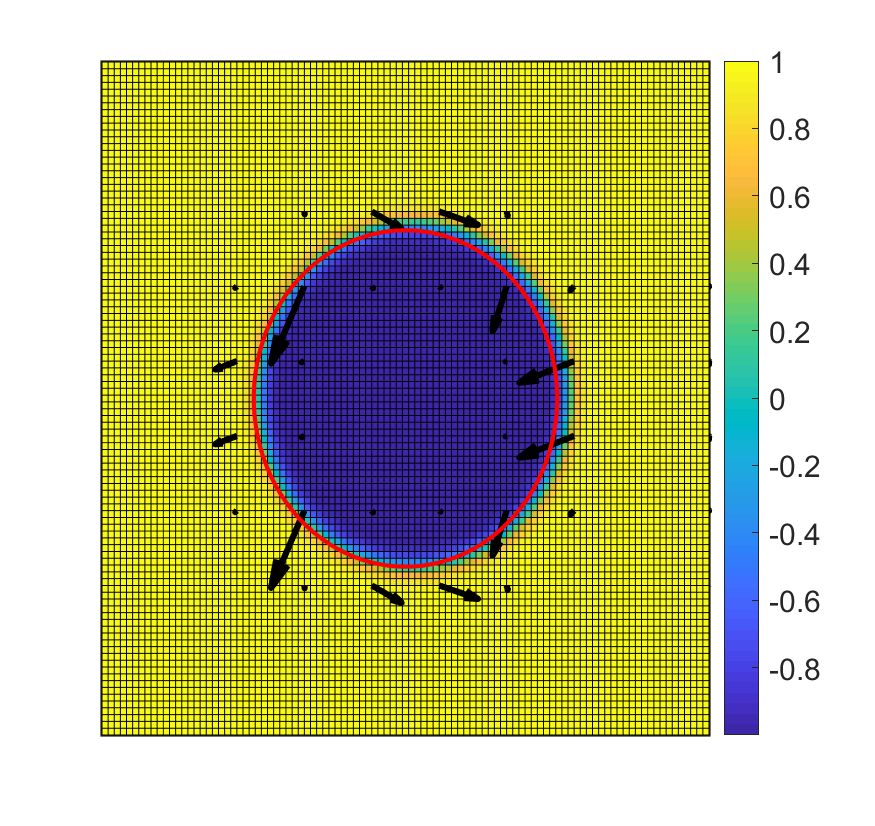}}
\caption{Droplet soliton given by Eqns.~\ref{e:dropletSoliton1} and~\ref{e:dropletSoliton2}.  Left image is a centered droplet soliton ($\boldsymbol\xi={\bf v}={\bf 0}$), right image is a non-centered droplet soliton ($|\boldsymbol\xi|,|{\bf v}|>0$).  False color represents out-of-plane component of $\hat{\bf m}$. Arrows represent magnitude and direction of in-plane component of $\hat{\bf m}$, which precesses as time $t$ advances.  The red circle is the boundary of the nanocontact.}
\label{f:dropletSoliton}
\end{figure}

Following Refs.~\cite{bookman_analytical_2013}, \cite{bookman_perturbation_2015}, and~\cite{wills_deterministic_2016}, we express the free-layer magnetization in spherical coordinates such that $\hatm=(\sin\Theta\cos\Phi,\sin\Theta\sin\Phi,\cos\Theta)$, with the droplet given approximately by
\begin{align}
&\cos\Theta = \tanh(\rho-1/\omega),\label{e:dropletSoliton1}\\
&\Phi = h_0t - \frac{{\bf v}\cdot\hat{\boldsymbol\rho}}{\omega^2}+\phi,\label{e:dropletSoliton2}
\end{align}
where $\phi = \omega t + \phi_0$ and it is assumed that the precessional frequency $\omega$ above the field-induced frequency $h_0$ is small, i.e.,
$0<\omega<0.25$, and the droplet velocity ${\bf v}$ is small, i.e., $|{\bf v}|\ll \omega$.
In this expression, $(\rho,\varphi)$ are polar coordinates centered at the droplet, related to the lab-frame coordinates through ${\bf x} = {\boldsymbol\xi} + \rho(\cos\varphi,\sin\varphi)$ where ${\boldsymbol\xi}_t={\bf v}$.  Figure~\ref{f:dropletSoliton} illustrates two examples of droplet solitons with the functional form given by~(\ref{e:dropletSoliton1}) and~(\ref{e:dropletSoliton2}).  These are described later in the text as centered (left) and non-centered (right) droplet solitons.

The droplet is thus characterized by six free parameters: azimuthal phase $\phi_0$, position ${\boldsymbol\xi}$, precessional frequency $\omega>0$, and velocity ${\bf v}$.  It was demonstrated in Ref.~\cite{bookman_perturbation_2015} using soliton perturbation theory in the limit of large nanocontact radius $\rho_*\gg 1$, weak damping and STT current $i = {\mathcal O}(\alpha)\ll 1$, and low temperature $\beta_0\ll 1$, that the droplet parameters evolve according to the SDEs
\begin{align}
&d\phi = \omega\,dt + \frac{i}{4\pi}\int_{x\leq\rho_*}\sech^2(\rho-1/\omega)\,d{\bf x}\,dt+dW_{\phi},\label{e:phiEqn}\\
&d\boldxi = {\bf v}\,dt + \frac{i\omega}{2\pi}\int_{x\leq \rho_*}\sech^2(\rho-1/\omega)\hat{\boldrho}\,d{\bf x}\,dt + d{\bf W}_{\xi},
\label{e:xiEqn}\\
&d\omega = \alpha\omega^2(\omega+h_0)\,dt-\frac{i\omega^3}{4\pi}\int_{x\leq \rho_*}\sech^2(\rho-1/\omega)\,d{\bf x}\,dt + dW_{\omega},\quad\mbox{and}
\label{e:omEqn}\\
&d{\bf v} = \alpha\omega(\omega+2h_0){\bf v}\,dt - \frac{i\omega^2}{2\pi}\int_{x\leq\rho_*}\left(\frac32{\bf v}-\frac{({\bf v}\cdot{\hat{\boldsymbol\varphi}})}{\rho\omega}
{\hat{\boldsymbol\varphi}}\right)\sech^2(\rho-1/\omega)\,d{\bf x}\,dt + d{\bf W}_v,\label{e:vEqn}
\end{align}
where the process ${\bf W} = (W_{\phi},{\bf W}_{\xi},W_{\omega},{\bf W}_v)$ satisfies
\be
{\mathbb E}[{\bf W}(t){\bf W}^\dag(t')] = \frac{\beta_0^2}{2\pi}\min(t,t')\begin{pmatrix}
\begin{matrix} v^2/4\omega+\omega/2 & {\bf v}^\dag/2 \\
{\bf v}/2 & \omega {\bf I}_{2\times2} \end{matrix} & {\bf 0}_{3\times 3} \\
{\bf 0}_{3\times 3} &
\begin{matrix}\omega^5/2 & \omega^4{\bf v}^T\\ \omega^4{\bf v} & {\boldsymbol \sigma}_v^2 \end{matrix}\end{pmatrix}
\ee
where
\be
{\bf\sigma}_v^2 = \begin{pmatrix}\omega^5+\omega^3(9v_x^2+v_y^2)/4 & 2\omega^3v_xv_y\\
2\omega^3v_xv_y & \omega^5+\omega^3(v_x^2+9v_y^2)/4\end{pmatrix}.
\ee

It should be pointed out that in deriving Eqn.~\ref{e:vEqn} from Eqn.~\ref{e:LL} (i.e., from the general modulation equation~(4.4) in Ref.~\cite{bookman_perturbation_2015}), the stochastic driving term is not strictly normalizable (i.e., does not have bounded variance) due to the $1/\rho$ contribution in the integrand near $\rho=0$ (the deterministic term does not suffer this difficulty since $1/\rho$ is integrable but not {\em square}-integrable).  Using the argument that the primary contribution from thermal fluctuations must come from values at the perimeter of the droplet $\rho\approx 1/\omega \gg 1$, however, we disregard the contribution near $\rho=0$ -- where the droplet is exponentially close to pointing down -- to provide a well-posed diffusion process.

Since no other parameters depend on $\phi$ and since spatial isotropy allows the coordinate frame to be aligned with ${\boldsymbol\xi}$ with nothing lost but an irrelevant angle, the full dynamics of Eqns.~\ref{e:phiEqn} through~\ref{e:vEqn} are captured by the four-dimensional system
\begin{align}
d\xi &= \left[\vpar + \frac{i\omega}{\pi}f_1(\xi,\omega)\right]\,dt+dW_{\xi},\label{e:xiSODE}\\
d\omega &= \left[\alpha\omega^2(\omega+h_0) - \frac{i\omega^3}{2\pi}f_2(\xi,\omega)\right]\,dt+dW_{\omega},\label{e:omSODE}\\
d\vpar &= \left[\alpha\omega(\omega+2h_0) - \frac{3i\omega^2}{2\pi}f_2(\xi,\omega) + \frac{i\omega}{\pi}f_3(\xi,\omega)\right]\vpar\,dt +dW_{\parallel},\quad\mbox{and}\label{e:parSODE}\\
d\vperp &= \left[\alpha\omega(\omega+2h_0) - \frac{3i\omega^2}{2\pi}f_2(\xi,\omega) + \frac{i\omega}{\pi}f_4(\xi,\omega)\right]\vperp\,dt +dW_{\perp},\label{e:perpSODE}
\end{align}
with $\vpar = {\bf v}\cdot{\hat{\boldsymbol\xi}}$, $\vperp = {\bf v}\cdot{\bf J}{\hat{\boldsymbol\xi}}$ where $J=\begin{psmallmatrix}0 & -1\\1 & 0\end{psmallmatrix}$, and
\be
{\mathbb E}[W(t)W^T(t')] = \frac{\beta_0^2}{2\pi}\min(t,t')\begin{pmatrix}\omega & 0 & {\bf 0}^T\\
0 & \omega^5/2 & \omega^4{\bf v}^T\\ {\bf 0} & \omega^4{\bf v} & {\boldsymbol \sigma}_v^2 \end{pmatrix}\label{e:4dDiffTensor}
\ee
where
\be
{\boldsymbol\sigma}_v^2 = \begin{pmatrix}\omega^5+\omega^3(9\vpar^2+\vperp^2)/4 & 2\omega^3\vpar\vperp\\
2\omega^3\vpar\vperp & \omega^5+\omega^3(\vpar^2+9\vperp^2)/4\end{pmatrix}
\ee
and ${\bf v}=(\vpar,\vperp)$.

The nonlocal terms in Eqns.~\ref{e:xiSODE} through~\ref{e:perpSODE} are given by
\begin{align}
&f_1(\xi,\omega) = \frac12\int_{x\leq\rho_*}\cos\varphi\,\sech^2(\rho-1/\omega)\,d{\bf x},\\
&f_2(\xi,\omega) = \frac12\int_{x\leq\rho_*}\sech^2(\rho-1/\omega)\,d{\bf x},\\
&f_3(\xi,\omega) = \frac12\int_{x\leq\rho_*}\frac{\sin^2\varphi}{\rho}\sech^2(\rho-1/\omega)\,d{\bf x},\quad\mbox{and}\\
&f_4(\xi,\omega) = \frac12\int_{x\leq\rho_*}\frac{\cos^2\varphi}{\rho}\sech^2(\rho-1/\omega)\,d{\bf x}.
\end{align}
The terms in~(\ref{e:phiEqn}) through~(\ref{e:vEqn}), and consequently~(\ref{e:xiSODE}) through~(\ref{e:perpSODE}),  proportional to $\alpha$ or $i$ correspond to contributions from magnetic damping or STT, respectively.

We study the above model using the parameters contained in Refs.~\cite{lendinez_observation_2015} and~\cite{lendinez_effect_2017}, where
$R_* = 75$~nm, $\mu_0 M_s = 0.95$~T, $\mu_0 H_k = 1.2$~T, $\lambda_{ex} = 5.3$~nm, $\mu_0 H_0 \in (0.2,1.4)$~T, $\delta=5$~nm, $\eta=0.26$, $\tau=0.13$~ns, $\alpha=0.03$, and $T=314$~K.  These parameters produce dimensionless parameters equal to $h_0 \in (0.8,5.6)$, $\rho_* = 7.3$, and $\beta^2=2.6\times 10^{-3}$.

\section{Large deviation theory}

Stochastic dynamical systems in the form of the Ito diffusion
\be
d{\bf u} = {\bf f}({\bf u})\,dt + \epsilon\beta({\bf u})\,d{\bf W}
\label{e:ItoDiffusion}
\ee
have long been the subject of study in physical systems where rare events are important phenomena to understand.  It is now well understood~\cite{freidlin_random_2012} that, in the limit as $\epsilon\rightarrow 0$, the probability of paths connecting any two states ${\bf u}_1$ and ${\bf u}_2$ approaches zero with a scaling that is exponential in $\epsilon^2$. The rate of this asymptotic decay is dictated by minimizers of the Wentzell-Freidlin action functional,
\be
S_T[{\bf u}(t)] = \int_0^T \frac12\left(\dot{\bf u}-{\bf f}({\bf u})\right)^T(\beta\left({\bf u})\beta({\bf u})^T\right)^{-1}\left(\dot{\bf u}-{\bf f}({\bf u})\right)\,dt\label{e:WFaction}
\ee
where the admissible set includes all absolutely continuous paths ${\bf u}(t)$ connecting ${\bf u}_1$ and ${\bf u}_2$.

In the case of an exit from the basin of attraction of a deterministically stable fixed point, the time $T$ over which the exit occurs clearly plays an important role in determining the probability of exit.  If $\Omega$ is the basin of attraction of a fixed point ${\bf u}_0$, then
\begin{align*}
&\lim_{\epsilon\rightarrow 0}\epsilon^2\log {\mathbb P}({\bf u}(T)\in\Omega) = -\min_{{\bf u}(T)\in \Omega\cup\partial\Omega}S_T[{\bf u}(t)],\\
&\lim_{\epsilon\rightarrow 0}\epsilon^2\log {\mathbb P}(t_E\leq T) = -\min_{{\bf u}(s)\notin\Omega,\, s\leq T}S_T[{\bf u}(t)],
\end{align*}
where $t_E = \min(t: {\bf u}(t)\notin\Omega)$ is defined as the first time of exit and $\partial\Omega$ is the boundary of $\Omega$.

Over arbitrary long time scales, exits occur with probability one.  The question is therefore not whether an exit will occur, but rather how long one should expect to wait before it does.  The mean time to exit (MTE) can be shown to increase exponentially as $\epsilon$ decreases, with a rate dictated by the action $S_T$ minimized over all possible transit times.  With the action left only as a function of state space, one can define a ``quasi-potential'' function that quantifies the MTE from ${\bf u}_0$ to any other state ${\bf u}$, i.e. $Q({\bf u}) = \inf_{T} S_T$, where $S_T$
is the minimizer of $S_T[{\bf u}(t)]$ over continuous paths ${\bf u}(t)$ satisfying ${\bf u}(0)={\bf u}_0$ and ${\bf u}(T)={\bf u}$.
In particular,
\be
\lim_{\epsilon\rightarrow0}\epsilon^2\log\mn[t_E] = \inf_{{\bf u}\in\partial\Omega}Q({\bf u}).
\label{e:MTE}
\ee
Under broad conditions, minima of the quasi-potential occur at 1-saddles, which represent the most likely points of exit from the basin of attraction of the stable fixed point.

Gradient flows (\ref{e:ItoDiffusion}) where ${\bf f}({\bf u}) = -\nabla V({\bf u})$ for some potential function $V({\bf u})$ represent a special case in the theory of large deviations.  In this case, it can be shown~\cite{freidlin_random_2012} that
the quasi-potential between a minimum of $V$ and the nearest saddle point is simply given by twice the difference in potential between the two points.  One-dimensional SDEs are automatically gradient flows since one can simply define
\be
V(u) = -\int_{u_0}^u f(y)\,dy.
\label{e:grad1d}
\ee

While the above consideration provides the exponential scaling law for transition probabilities and mean times to exit, a more detailed analysis is required~\cite{gardiner_stochastic_2009,forgoston_primer_2018,matkowsky_exit_1977} to improve on this estimate by providing the prefactor of the exponential, for example.  In this work we discuss only the rate function, leaving a more accurate approximation for further study.

\section{Deterministic dynamics}
\label{s:detDyn}

In the four-dimensional system described by Eqns.~\ref{e:xiSODE} through~\ref{e:perpSODE}, the first three parameters only depend on $\vperp$ through the diffusion tensor.  A complete description of the deterministic dynamics is therefore afforded by the reduced system
\begin{align}
\dot\xi &= \vpar + {i\omega}{\pi}f_1,\label{e:xiDet}\\
\dot\omega &= \alpha\omega^2(\omega+h_0)-\frac{i\omega^3}{2\pi}f_2,\label{e:omDet}\\
\dot\vpar &= \left[\alpha\omega(\omega+2h_0) - \frac{3i\omega^2}{2\pi}f_2 + \frac{i\omega}{\pi}f_3\right]\vpar.\label{e:vDet}
\end{align}

\begin{figure}
\centerline{\includegraphics[width=0.5\textwidth]{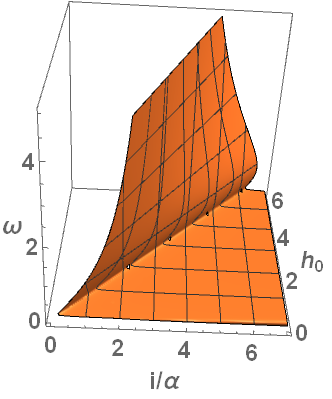}}
\caption{Zero-level surface of $\Gamma(\omega;i/\alpha,h_0)$ for $\rho_*=7.3$.}
\label{f:levelSurf}
\end{figure}

The case $\vpar = 0$ has been thoroughly studied in Ref.~\cite{bookman_analytical_2013}, where it was shown that multiple fixed points are possible with $\xi=\vpar=0$ depending on the number of roots of
\be
\Gamma(\omega) := \alpha(\omega+h_0) - \frac12i\omega\left[\rho_*\tanh\left(\rho_*-\frac1{\omega}\right)-\ln\cosh\left(\rho_*-\frac1{\omega}\right) +\ln\cosh\frac1\omega\right],
\label{e:heqn}
\ee
where we note that $\Gamma(0^+) = \alpha h_0>0$.  Figure~\ref{f:levelSurf} illustrates the zero-level surface of $\Gamma(\omega)$ for different values of the applied field $h_0$ and scaled STT current $i/\alpha$ for $\rho_*=7.3$.  Considering for a moment the 2d system where we restrict $\vpar=0$ and fix a small applied field $h_0$, a saddle-node bifurcation occurs for $i/\alpha\approx h_0$ and the saddle approaches infinity as $i/\alpha$ approaches $2/\ln 2$.  For larger values of applied field, another stable fixed point bifurcates from infinity either before or after the original saddle-node bifurcation to collide with and annihilate the saddle as $i/\alpha$ is increased.
Since $\sgn(f_1) = -\sgn(\xi)$ (see Appendix), fixed points satisfy $\vpar=0$ if and only if $\xi=0$.  {\em Centered} droplet solitons, depicted in the left image of Fig.~\ref{f:dropletSoliton}, satisfy $\xi=\vpar=0$ while {\em non-centered} droplet solitons, depicted in the right image, satisfy $\xi\neq 0$ and $\vpar\neq 0$.
%

The full stability of the centered fixed points is dictated by the Jacobian evaluated at $(0,\omega_*,0)$,
\be
J = \begin{pmatrix} \lambda_1 & 0 & 1\\
0 & \lambda_2 & 0\\
0 & 0 & \lambda_3\end{pmatrix},
\ee
where
\begin{align}
&\lambda_1 = -\frac12\rho_*i\omega\,\sech^2(\rho_*-1/\omega),\\
&\lambda_2 = \lambda_1-\alpha h_0\omega+\frac12i\omega\left(\tanh(\rho_*-1/\omega)+\tanh(1/\omega)\right),\\
&\lambda_3 = \lambda_2-\lambda_1-2\alpha\omega^2.\label{e:lam3}
\end{align}
Bifurcation to the drift instability described in Ref.~\cite{wills_deterministic_2016} occurs when $i/\alpha$ is sufficiently large to force $\lambda_3>0$, which occurs when
\be
\frac{i}{\alpha} = \frac{2(2\omega_*+h_0)}{\tanh(\rho_*-1/\omega_*)+\tanh(1/\omega_*)}.
\label{e:driftBifCond}
\ee
The tangent direction of this unstable manifold is
\be
\psi_3 = \begin{pmatrix} 1 & 0 & \frac12\rho_*i\omega_*\sech^2(\rho_*-1/\omega_*)\end{pmatrix},
\ee
providing an indication of the ejection mechanism associated with the drift instability.

\begin{figure}
\centerline{\includegraphics[width=\textwidth]{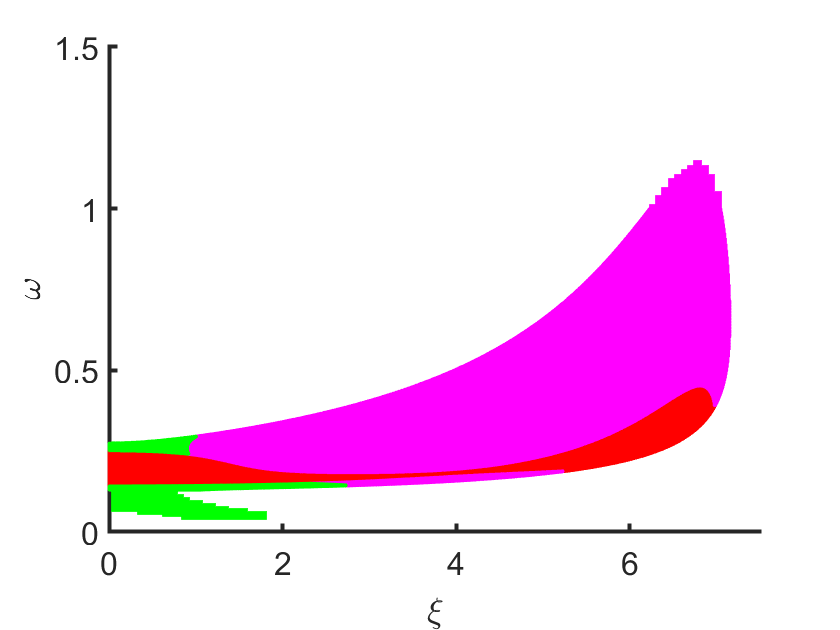}}
\caption{Phase diagram of non-centered fixed points plotted vs.\ $\xi$ and $\omega$ and color-coded by stability (green is stable; red and magenta correspond to one- and two-dimensional unstable manifolds, respectively).  Whitespace region corresponds to unphysical (i.e., negative) values of $i/\alpha$ or $h_0$.  Slight pixelation for small and large values of $\omega$ reflect lower resolution of computational grid in those regions.}
\label{f:NonCenteredFPvsXiOm}
\end{figure}

\begin{figure}
\centerline{\includegraphics[width=\textwidth]{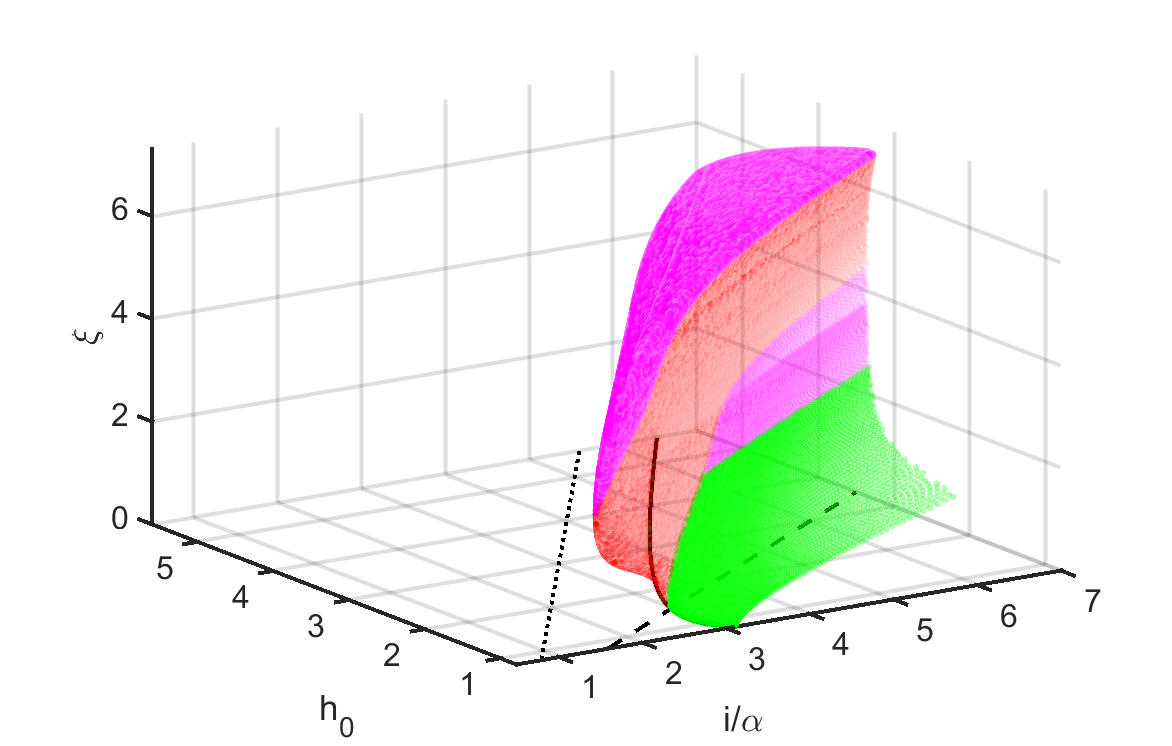}}
\caption{Families of non-centered fixed points color-coded by stability (green is stable; red and magenta correspond to one- and two-dimensional unstable manifolds, respectively) for $\rho_*=7.3$.  Each bifurcates from the drift instability identified in~\cite{wills_deterministic_2016} (solid black) at $\xi=0$ and moves with increasing position $\xi$ to accumulate in the region (demarcated by dotted and dashed black lines) identified asymptotically in Eqn.~(\ref{e:FPcond}).  Multiple fixed points can exist for each parameter pair $(i/\alpha,h_0)$, including in the region where $(0,\omega_*,0)$ is stable.}
\label{f:nonzeroFPxi}
\end{figure}

\begin{figure}
\centerline{\includegraphics[width=\textwidth]{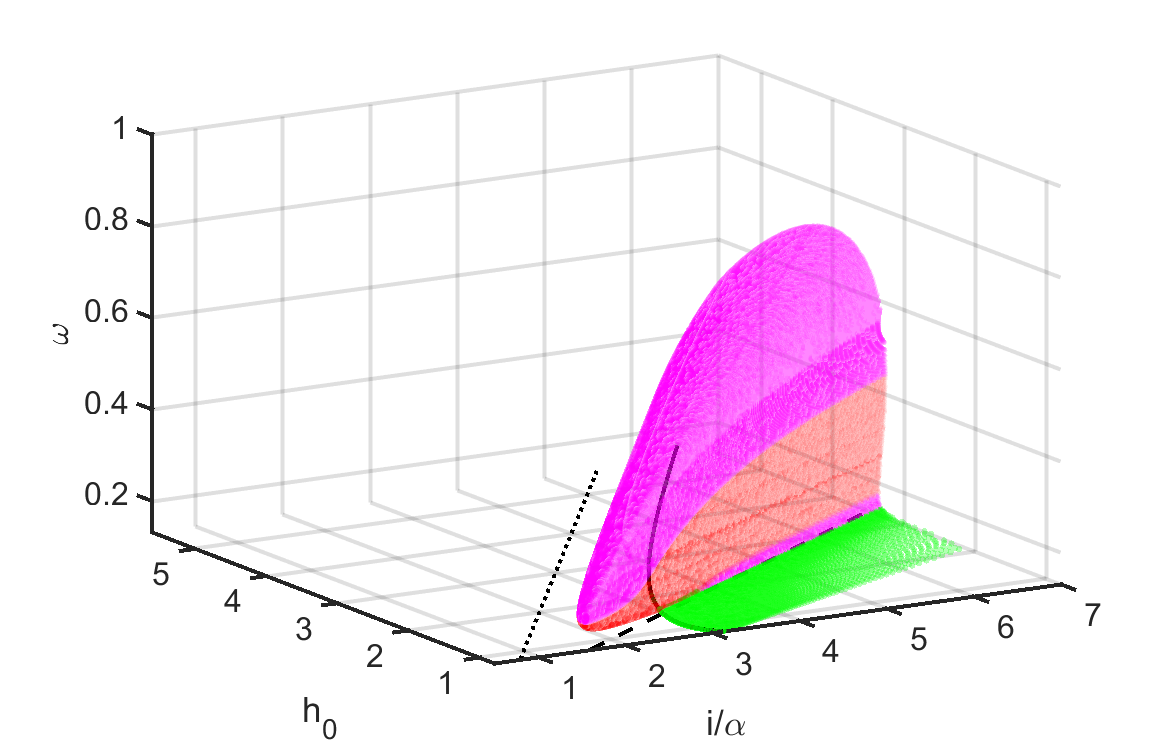}}
\caption{Graph of $\omega$-values associated with the fixed points depicted in Fig.~\ref{f:nonzeroFPxi}, emerging from $\omega=\omega_*$ at the drift bifurcation curve and accumulating in the region identified in Eqn.~\ref{e:FPcond}.  Color coding is the same as in Figs.~\ref{f:NonCenteredFPvsXiOm} and~\ref{f:nonzeroFPxi}.}
\label{f:nonzeroFPomega}
\end{figure}

\begin{figure}
\centerline{\includegraphics[width=\textwidth]{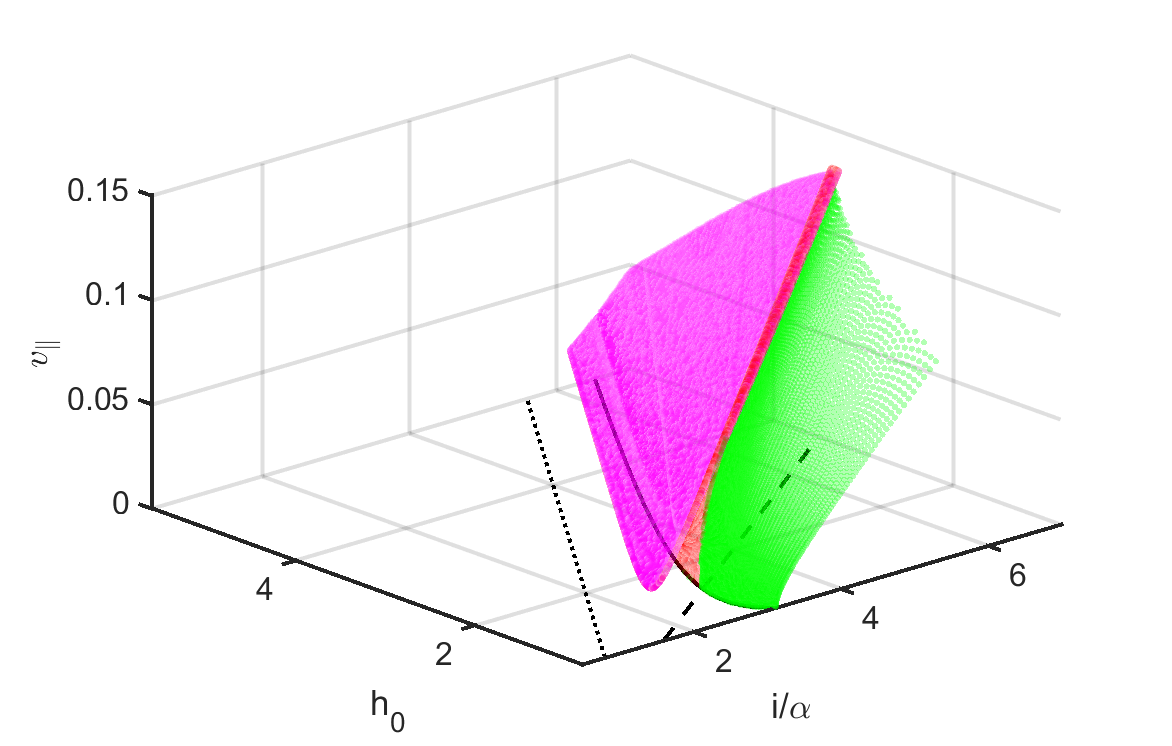}}
\caption{Graph of $\vpar$-values associated with the fixed points depicted in Fig.~\ref{f:nonzeroFPxi}, emerging from $\vpar=0$ at the drift bifurcation curve and accumulating in the region identified in Eqn.~\ref{e:FPcond}.  Color coding is the same as in Figs.~\ref{f:NonCenteredFPvsXiOm} and~\ref{f:nonzeroFPxi}.}
\label{f:nonzeroFPv}
\end{figure}

Analysis of non-centered fixed points with $\vpar\neq 0$ ($\xi\neq 0$) is greatly simplified by the fact that they satisfy a two-dimensional algebraic system
\begin{align}
& \frac{\alpha}{i}(\omega+h_0) - \frac{\omega}{2\pi}f_2(\xi,\omega)=0,\label{e:FP1}\\
& \frac{\alpha}{i}(2\omega+h_0) - \frac1{\pi}f_3(\xi,\omega)=0
\label{e:FP2}
\end{align}
with the velocity then provided by
\begin{align}
\vpar = -\frac{i\omega}{\pi}f_1(\xi,\omega).
\end{align}
Rather than computing fixed points $(\xi,\omega,\vpar)$ associated with relevant parameter combinations $(\alpha,i,h_0)$, it is convenient to compute the unique parameter sets $(i/\alpha,h_0)$ associated with each potential fixed point in the $(\xi,\omega)$ plane and verify that the parameters are within the appropriate asymptotic regime for perturbation theory to be applicable.  Thus, the subset of non-centered fixed points plotted in Fig.~\ref{f:NonCenteredFPvsXiOm} versus $\xi$ and $\omega$ that correspond to physically relevant values of $i/\alpha$ and $h_0$ result in
 Figs.~\ref{f:nonzeroFPxi}, \ref{f:nonzeroFPomega}, and~\ref{f:nonzeroFPv}.
 Each of these images is in fact a surface resolved through a set of points using the method described above, where the uniform grid defined in the $(\xi,\omega)$ plane is plainly visible.  The color-coded points indicate stability, with green denoting stable fixed points, red denoting fixed points with a one-dimensional unstable manifold (referred to here as 1-saddles), and magenta denoting fixed points with a two-dimensional unstable manifold.  The $(i/\alpha,h_0)$-plane at the base of each graph illustrates the dashed lines $i=\alpha h_0$, which provides an approximate lower bound for existence of the centered equilibrium, and the value $i=2\alpha h_0$.  The solid line is the drift bifurcation curve, computable using Eqn.~\ref{e:driftBifCond}, from which it can be seen that all non-centered fixed points emerge.
At small values of $h_0$, the bifurcation is supercritical in $i/\alpha$ such that the centered equilibrium immediately transfers its stability to the new non-centered equilibrium.  For larger values of $h_0$, however, the bifurcation is subcritical in $i/\alpha$, with a sheet of saddles having only one unstable direction in close proximity to the stable centered equilibria.  These present the primary ejection mechanism when thermal effects are included, and will be discussed in the next section.

\begin{figure}
\centerline{\includegraphics[width=0.5\textwidth]{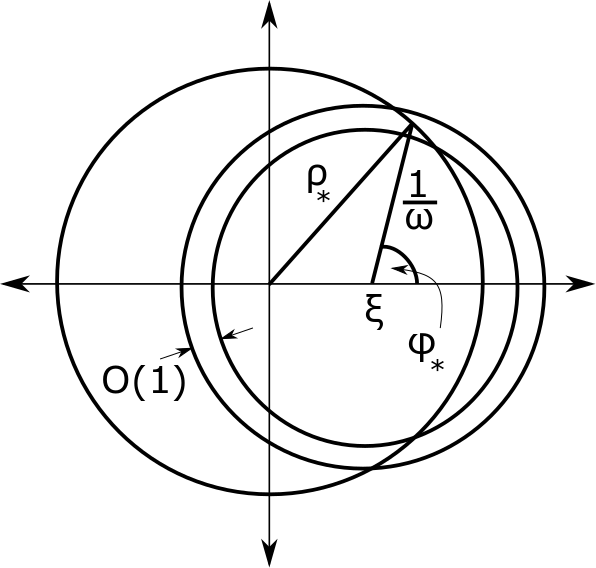}}
\caption{Illustration of geometry used in asymptotic approximation of non-centered fixed point.  Soliton-nanocontact interaction only occurs in the portion of the annular region (droplet soliton) that lies inside the zero-centered circle (nanocontact).}
\label{f:asympFig}
\end{figure}
An alternative to the strategy above for solving Eqns.~\ref{e:FP1} and~\ref{e:FP2} that provides additional insight is to consider them asymptotically for large $\rho_*$ and small $\omega$ (i.e., the regime of validity for Eqns.~(\ref{e:xiDet}) through~(\ref{e:vDet})).  We observe in that case that the support of each integral lies in the intersection between the circular disk of radius $\rho_*$ and a ``washer'' with ${\mathcal O}(1)$ width and central radius $1/\omega$, with the disk and washer centers separated by $\xi$, as depicted in Fig.~\ref{f:asympFig}.  The two points of intersection lie at angles $\pm\varphi_*$ satisfying
\[
\cos(\varphi_*)= \frac{\omega}{2\xi}\left(\rho_*^2-\frac1{\omega^2}-\xi^2\right),
\]
so that by moving into droplet-centered coordinates, we have
\begin{align}
\int_{x\leq \rho_*}\sech^2(\rho-1/\omega)\,d{\bf u}&\approx \int_{\varphi_*}^{2\pi-\varphi_*}\int_{-\infty}^{\infty}\sech^2(\rho-1/\omega)\rho\,d\rho\,d\varphi\\
&\approx \frac4{\omega}(\pi-\varphi_*),
\end{align}
where $0\leq\varphi_*<\pi$ by construction.  The implicit assumption in this construction is that the disk and annulus have nontrivial intersection, which approximately equates to the condition
\be
|\xi-\rho_*| < 1/\omega < \xi + \rho_*.
\ee
Treating the remaining integrals similarly, we have that
\begin{align}
\pi\frac{\alpha}{i}(\omega+h_0)\approx \pi-\varphi_*,\label{e:asymp1}\\
\pi\frac{\alpha}{i}(2\omega+h_0)\approx \pi-\varphi_* + \frac12\sin 2\varphi_*.\label{e:asymp2}
\end{align}
Equating expressions for $\omega$ thus gives
\be
\sin 2\varphi_* = 2(\pi-\varphi_* - \pi\alpha h_0/i).
\ee
From Eqn.~\ref{e:asymp1}, we require that $\pi-\varphi_* > \pi\alpha h_0/i$, so that
\be
0\leq \varphi_* \leq \pi/2,
\ee
which confines the droplet center to within the nanocontact (see Fig.~\ref{f:asympFig}).  Thus
\be
\alpha h_0 < i < 2\alpha h_0
\label{e:FPcond}
\ee
for existence of this fixed point.  Figures~\ref{f:nonzeroFPxi} through~\ref{f:nonzeroFPv} clearly demonstrate that the surface of fixed points emerging from the drift bifurcation accumulates within this region.  Moreover, it is only within this region that the drift bifurcation is subcritical, such that the region where a stable fixed point coexists with a saddle point associated with the drift mode is the intersection of~(\ref{e:FPcond}) and the half-plane bounded by~(\ref{e:driftBifCond}).

The unique solution for $\varphi_*$ produces non-centered fixed points with approximate values
\begin{align}
&\omega_v \approx \frac{i}{\alpha}\left(1-\frac{\varphi_*}{\pi}\right) - h_0\quad\mbox{and}\\
&\xi_v \approx \sqrt{\rho_*^2-\frac{\sin^2\varphi_*}{\omega_v^2}}-\frac{\cos\varphi_*}{\omega_v}.\label{e:xiAsympt}
\end{align}
Both of these quantities are positive.  The associated parallel velocity is provided by
\be
v_v \approx \frac{2i}{\pi}\sin\varphi_*,
\ee
which is well-approximated near the critical line $i=\alpha h_0$ by
\be
v_v \approx i\left(\frac{i}{\alpha h_0}-1\right).
\label{e:vAsymp}
\ee
The perpendicular velocity $\vperp$ associated with non-centered fixed points is $\vperp=0$.  Hence, for each $(\omega_v,\xi_v,v_v)$ fixed point, there is a corresponding collection of fixed points for the original six-dimensional dynamical system~(\ref{e:phiEqn}-\ref{e:vEqn}) with droplet positions ${\boldsymbol\xi}$ on the circle $|{\boldsymbol\xi}|=\xi_v$.
Having identified all centered and non-centered fixed points, we now turn to the calculation of the action of paths connecting them.

\section{Action and optimal paths}

In the limit of vanishing noise strength, exits from the basin of attraction of a stable fixed point collapse at an exponentially growing probability onto paths that lead to the nearest 1-saddle, where proximity is defined using a metric weighted by the diffusion tensor.  In the four-dimensional system given by Eqns.~\ref{e:xiSODE} through~\ref{e:perpSODE}, there are two candidates for the saddle that represent the most likely exit point, either a decay mode or a drift mode.  We study these in turn to identify their associated limiting exit rates.

The first saddle is that identified at $\xi=v=0$ for values of $i/\alpha$ that lie between the saddle-node bifurcation occurring at approximately $i/\alpha=h_0$ and the value of $i/\alpha$ where the saddle either runs off to infinity (for small values of $h_0$) or is destroyed in a saddle-node bifurcation with another stable fixed point (for sufficiently large values of $h_0$).  Labelling the centered stable fixed point by $(0,\omega_*,0)$ and the saddle by $(0,\omega_s,0)$, we postulate based on the form of the action~(\ref{e:WFaction}) and numerical results that optimal paths connecting these two fixed points with optimal action $S$ satisfy $\xi=|{\bf v}|=0$.  Dependence of the diffusion tensor on the state variables precludes a definite statement that this is true, but we have observed it in all cases we have computed.  Since this exit mechanism occurs strictly via changes in $\omega$, thereby mimicking the impact of damping on solitons~\cite{bookman_analytical_2013}, we refer to it as a decay mode of exit.

\begin{figure}
\centerline{\includegraphics[width=0.7\textwidth]{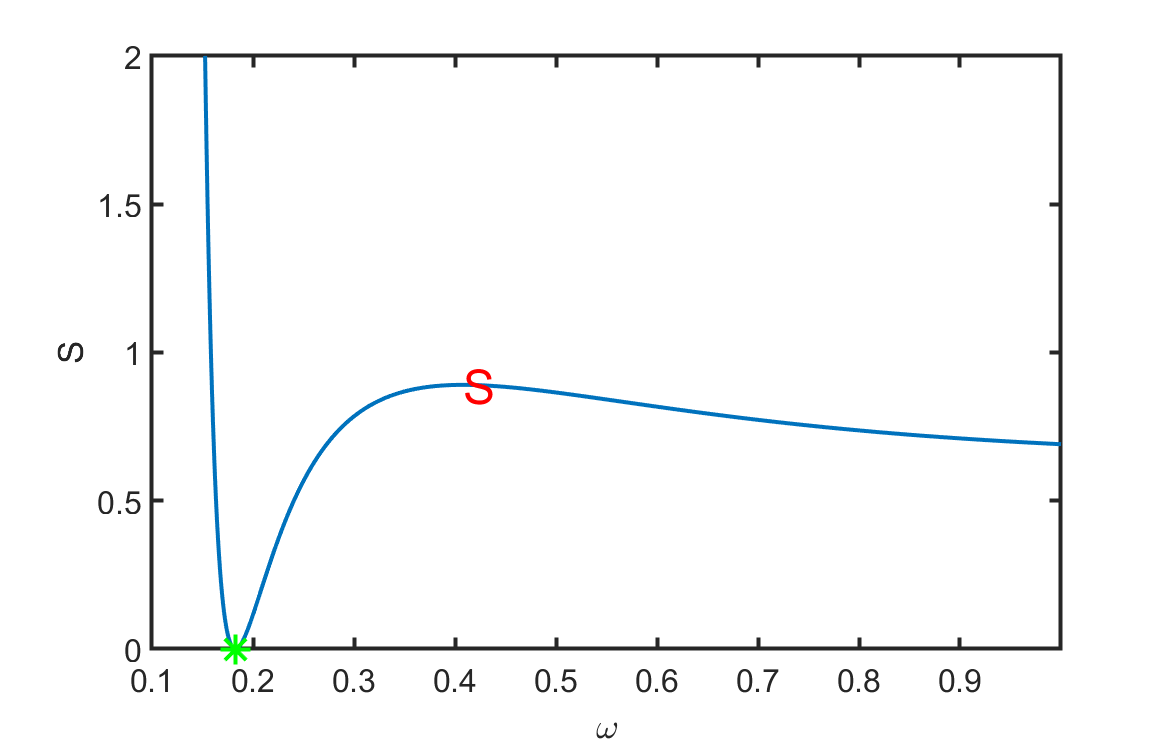}}
\caption{Quasipotential given in Eqn.~\ref{e:quasipotential} for $\rho_*=7.3$, $i/\alpha=6$, $h_0=5.6$.  Green star and red ``S'' denote stable fixed point and saddle, respectively.  $Q$ blows up as $\omega\rightarrow 0^+$.}
\label{f:quasipotential}
\end{figure}

If we assume this to be true, the action collapses to
\be
S = \int_{-\infty}^{\infty} L\,dt\qquad\mbox{with}\qquad L = \frac{2\pi}{\omega^5}[\dot{\omega}-\omega^2\Gamma(\omega)]^2.
\ee
where, by construction, $\Gamma(\omega)\leq 0$ if $\omega_*\leq\omega\leq\omega_s$ and equality holds only at the endpoints.
As explained on p.~\pageref{e:grad1d}, one-dimensional flows are trivially gradient flows and it is immediate that optimal paths satisfy
\be
\dot\omega = -\omega^2\Gamma(\omega),
\ee
which integrates to give the Wentzell-Freidlin action
\be
S = Q(\omega_s)-Q(\omega_*)
\label{e:quasi}
\ee
where $Q$ is the quasi-potential defined by
\begin{align}
Q(\omega) &= -8\pi\int \frac{\Gamma(\omega)}{\omega^3}\,d\omega\nonumber\\
&=4\pi\left(\frac{h_0\alpha-i}{\omega^2} + \frac{2\alpha}{\omega} + \rho_*i\ln\frac{\cosh(\rho_*-1/\omega)}{\cosh\rho_*}
+\frac{\rho_*i}{\omega}\right.\nonumber\\
& \left.-\frac{i}2\left[\Li_2(-e^{-2(\rho_*-1/\omega)})+\Li_2(-e^{-2/\omega})-\Li_2(-e^{-2\rho_*})+\frac{\pi^2}{12}\right]\right)
\label{e:quasipotential}
\end{align}
with $\Li_2(z)$ the dilogarithm function defined by
\be
\Li_2(z) = -\int_0^z\ln(1-t)\frac{dt}{t}.
\ee
The quasi-potential~(\ref{e:quasipotential}) has been defined such that $\lim_{\omega\rightarrow\infty}Q(\omega)=0$.  As explained earlier, depending on the parameters $h_0$ and $i/\alpha$, the saddle point $(0,\omega_s,0)$ may not exist.  In these cases, we identify an exit via the decay mode with $\omega\rightarrow\infty$, such that the associated Wentzell-Freidlin action simplifies to $S=-Q(\omega_*)$.
As illustrated in Fig.~\ref{f:quasipotential}, expression~(\ref{e:quasipotential}) for the quasipotential $Q$ blows up as $\omega\rightarrow 0$.  This is expected since the noise strength goes to zero as $\omega\rightarrow 0$.
%
%
%
%
\begin{figure}
\centerline{\includegraphics[width=0.8\textwidth]{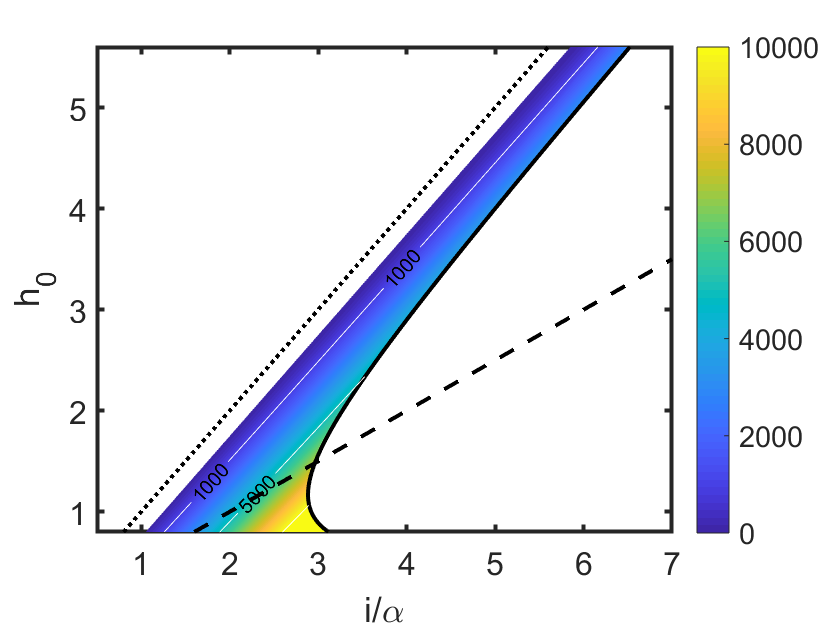}}
\caption{Log of MTE (in dimensionless time units) corresponding to annihilation of the droplet soliton via noise-induced damping.  Except for parameters close to the dotted line approximating the existence boundary for the stable centered fixed point, the MTE from this damping mechanism is effectively infinite.}
\label{f:MTEviaDampedMode}
\end{figure}
Figure~\ref{f:MTEviaDampedMode} depicts the Wentzell-Freidlin action~(\ref{e:quasi}) for different relative STT currents $i/\alpha$ and different applied magnetic fields $h_0$.  Moving from the dotted line, which provides a lower bound on values of $i/\alpha$ for which the centered droplet exists, toward the solid line, at which the centered droplet loses stability via the drift bifurcation,
the MTE is observed to increase very quickly, becoming effectively infinite for parameters not within a very small margin of the dotted line.

\begin{figure}
\centerline{\includegraphics[width=0.5\textwidth]{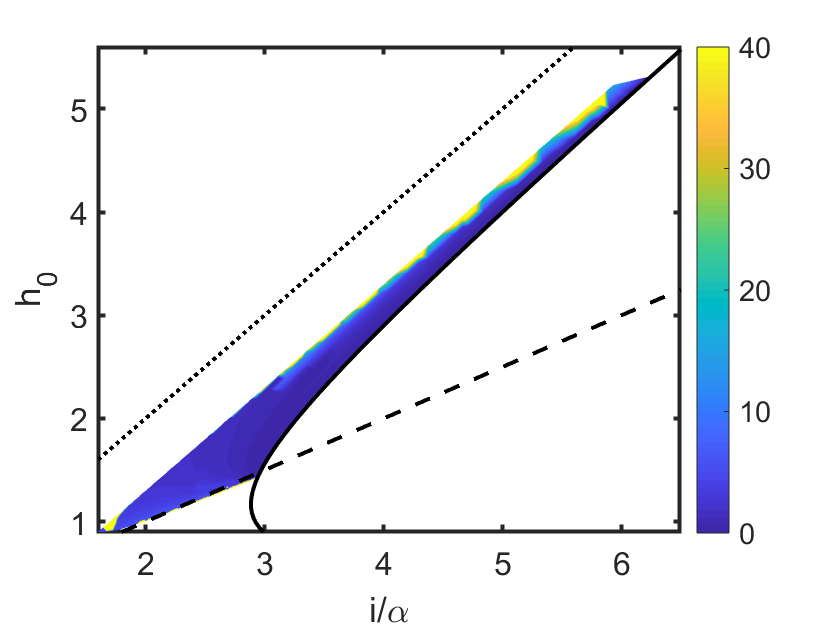}
\includegraphics[width=0.5\textwidth]{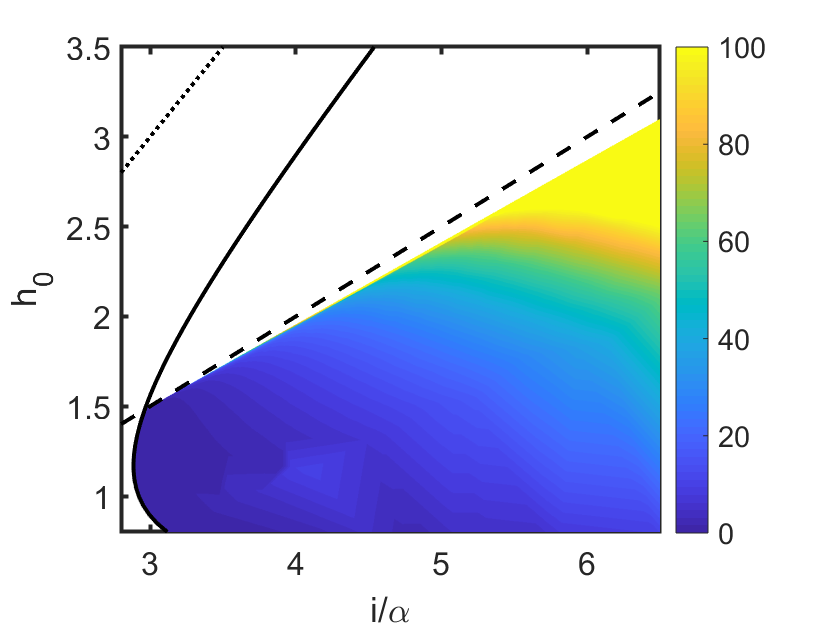}}
\caption{Log of MTE (in dimensionless time units) corresponding to annihilation of the droplet soliton via the drift instability.  Left: MTE from stable centered fixed points through non-centered saddle points.  Note that the simultaneous existence (i.e., for the same values of $i/\alpha$ and $h_0$) of these two fixed points dictates the parameters for which this calculation is relevant (see Fig.~\ref{f:nonzeroFPxi}).  Right: MTE from stable non-centered fixed points through centered saddle points.  This calculation is relevant for a different region of parameter space.  Note that the exit times associated with this droplet annihilation mechanism are much lower than the exit times shown in Fig.~\ref{f:MTEviaDampedMode}, and these events are therefore much more likely to be observed on any finite time scale.}
\label{f:MTEviaDriftMode}
\end{figure}

Computing the minimum action between the stable fixed point and the saddle with nonzero velocity requires, in principle, consideration of the four-dimensional system given by Eqns.~\ref{e:xiSODE} through~\ref{e:perpSODE} due to the dependence of the diffusion tensor given by Eqn.~\ref{e:4dDiffTensor} on $\vperp$.  The contribution from nonzero $\vperp$ is expected to be negligible because $\vperp=0$ at all fixed points, however.  For simplicity, we therefore restrict our computations to the minimizers of the action restricted to three-dimensional dynamics, giving
\be
S = \int_{-\infty}^{\infty} \frac12\left(\dot{\bf u}-{\bf f}({\bf u})\right)^T(\beta\left({\bf u})\beta({\bf u})^T\right)^{-1}\left(\dot{\bf u}-{\bf f}({\bf u})\right)\,dt
\label{e:Sagain}
\ee
where ${\bf u} = (\xi,\omega,\vpar)$,
\be
{\bf f} =  \begin{pmatrix} \vpar + \frac{i\omega}{\pi}f_1(\xi,\omega)\\
\alpha\omega^2(\omega+h_0) - \frac{i\omega^3}{2\pi}f_2(\xi,\omega)\\
\left[\alpha\omega(\omega+2h_0) - \frac{3i\omega^2}{2\pi}f_2(\xi,\omega) + \frac{i\omega}{\pi}f_3(\xi,\omega)\right]\vpar
\end{pmatrix}.
\ee
The diffusion tensor appearing in Eqn.~\ref{e:Sagain} is
\be
\beta\beta^T = \frac1{2\pi}\begin{pmatrix}\omega & 0 & 0\\
0 & \omega^5/2 & \omega^4\vpar\\ 0 & \omega^4\vpar &  \omega^5+\frac9{4}\omega^3\vpar^2 \end{pmatrix}.\label{e:3dDiffTensor}
\ee
The drift ${\bf f}({\bf u})$ is not, in general, either gradient or trivially non-gradient and the action must therefore be computed numerically.  The left image in Fig.~\ref{f:MTEviaDriftMode} depicts the approximate MTE obtained by using~(\ref{e:MTE}) with the action computed using the generalized minimum action method (or GMAM, see Ref.~\cite{heymann_geometric_2008} for details) at parameter values where the base stationary state is stable and coexists with the saddle corresponding to the drift mode.  As explained on p.~\pageref{e:FPcond} and observed in Figs.~\ref{f:nonzeroFPxi} through~\ref{f:nonzeroFPv}, for a given nanocontact radius $\rho_*$, this is approximately the set of current values $h_0<i/\alpha<2h_0$.
The right image in Fig.~\ref{f:MTEviaDriftMode} is the approximate MTE associated with leaving the basin of attraction of stable non-centered fixed points through the now-unstable centered fixed point.

In the parameter regime operable in Fig.~\ref{f:MTEviaDampedMode} and the left image of Fig.~\ref{f:MTEviaDriftMode}, the droplet soliton is potentially destroyed via two candidate noise-related mechanisms.  The first is the drift mechanism described above, where the position and precessional frequency are driven by noise to the point where the droplet loses stability to a transient traveling droplet which is then ejected from the nanocontact.  The MTE associated with this mechanism is plotted on the left in Fig.~\ref{f:MTEviaDriftMode}.
The second is the decay mode described earlier, whereby an improbable accumulation of fluctuations that drive its precessional frequency sufficiently high as to fall into a regime where damping is no longer counterbalanced by the current passing through the nanocontact.  The stochastic dynamics of this mechanism has an associated MTE depicted in Fig.~\ref{f:MTEviaDampedMode}.
%
%
It is clear from this comparison that the action associated with the drift mode is substantially smaller for all parameters investigated here, indicating that this is by far the most dominant mechanism for exit of a centered droplet soliton over the range of values of magnetic field $h_0$ and rescaled current $i/\alpha$ for which the original fixed point is stable and the saddle associated with the drift mode exists and has a one-dimensional unstable manifold.  As discussed earlier, this is true for a subset of rescaled current values satisfying $i/\alpha < 2h_0$.  For large current but small external field $h_0$, where the non-centered fixed points are stable, it appears that the only exit mechanism for those non-centered droplets is through the centered saddle point with a MTE somewhat higher than the exit times associated with annihilation of centered droplets; see Fig.~\ref{f:MTEviaDriftMode}, right for the MTE associated with this ejection mechanism.

\section{Conclusions}

Perturbation theory for a droplet soliton that experiences uniform
magnetic damping, local spin transfer torque via a nanocontact, a
vertical applied field, and thermal fluctuations, results in a
six-dimensional system of stochastic ordinary differential equations.
In this paper, we have reduced this system to an equivalent---in the
deterministic case---or approximately equivalent---in the stochastic
case---three-dimensional system that is amenable to detailed analysis.
We completely determined the nature of the fixed points for this dynamical
system.  One family of fixed points, called nanocontact-centered, exist and are deterministically stable in a restricted regime of current and applied field.  These fixed points represent dissipative droplet solitons centered within the nanocontact and precessing at a fixed frequency.
In addition to previously studied nanocontact-centered fixed
points, we also identify a new class of non-centered fixed points.
These latter fixed points correspond to droplet solitons
whose centers are displaced relative to the center of the nanocontact, with the resulting force balanced by an anisotropy in the soliton's in-plane magnetization.  These non-centered fixed points are stable in an appropriate parameter regime, but we have not found any parameters at which the centered and non-centered droplet solitons are simultaneously stable.
By combining our fixed point analysis with large deviation theory, we
identify the mostly likely path through phase space along which a
droplet subject to weak thermal noise is expected to decay.  A direct
computation of the Wentzell-Freidlin action shows that the most likely
decay process for stochastically perturbed, centered droplet fixed
points is to drift out of the nanocontact.  The computed action also
provides an estimate for the rate of droplet expulsion from the
nanocontact, which sets an effective lower bound for droplet stability
in these devices at finite temperature.  Based on the parameter values in Refs.~\cite{lendinez_effect_2017} and~\cite{lendinez_observation_2015}, with a time scale $\tau=0.13$~ns, the left image of Fig.~\ref{f:MTEviaDriftMode} yields an estimated droplet ejection rate of
$1/(\tau e^6) \approx 19~\mathrm{MHz}$ at room temperature.
This rate is an order of magnitude lower than the low-frequency observations in experiments that have been in the 100-500~MHz range.  It is important to note that estimates obtained by computing just the Wentzell-Freidlin action~(\ref{e:WFaction}) provide the scaling law of the MTE in the limit of small noise strength but neglect prefactors that can be large.

\bibliography{magneticSolitons,magneticSolitons2}

\end{document}